\input harvmac.tex
% \draftmode
%%%%%%%%%%%  Math-style letters   %%%%%%%%%
\font\cmss=cmss10 \font\cmsss=cmss10 at 7pt

\def\Z{\relax\ifmmode\mathchoice
{\hbox{\cmss Z\kern-.4em Z}}{\hbox{\cmss Z\kern-.4em Z}}
{\lower.9pt\hbox{\cmsss Z\kern-.4em Z}}
{\lower1.2pt\hbox{\cmsss Z\kern-.4em Z}}\else{\cmss Z\kern-.4em Z}\fi}

\def\hat{\widehat}
\def\bar{\overline}
%%%%%%%%%%%%%%%%%%%%%%%%%%  FIGURES   %%%%%%%%%%%%%%%%%%%%%%%%%%%%%%%

\let\includefigures=\iftrue
\newfam\black
\includefigures
\input epsf
\def\figin{\epsfcheck\figin}\def\figins{\epsfcheck\figins}
\def\epsfcheck{\ifx\epsfbox\UnDeFiNeD
\message{(NO epsf.tex, FIGURES WILL BE IGNORED)}
\gdef\figin##1{\vskip2in}\gdef\figins##1{\hskip.5in}% blank space instead
\else\message{(FIGURES WILL BE INCLUDED)}%
\gdef\figin##1{##1}\gdef\figins##1{##1}\fi}
\def\DefWarn#1{}
\def\figinsert{\goodbreak\midinsert}
\def\ifig#1#2#3{\DefWarn#1\xdef#1{fig.~\the\figno}
\writedef{#1\leftbracket fig.\noexpand~\the\figno}%
\figinsert\figin{\centerline{#3}}\medskip\centerline{\vbox{\baselineskip12pt
\advance\hsize by -1truein\noindent\footnotefont{\bf Fig.~\the\figno:} #2}}
\bigskip\endinsert\global\advance\figno by1}
%%%
\else
\def\ifig#1#2#3{\xdef#1{fig.~\the\figno}
\writedef{#1\leftbracket fig.\noexpand~\the\figno}%
%\figinsert\figin{\centerline{#3}}\medskip\centerline{\vbox{\baselineskip12pt
%\advance\hsize by -1truein\noindent\footnotefont{\bf Fig.~\the\figno:} #2}}
%\bigskip\endinsert
\global\advance\figno by1}
\fi

%%%%%%%%%%%%% Calligraphic letters  %%%%%%

\def\CF {{\cal F}}

\def\CK {{\cal K}}

\def\CM {{\cal M}}
\def\CN {{\cal N}}

\def\CV {{\cal V}}

\lref\ADS{J.M.~ Maldacena, "The Large N Limit of Superconformal
Field Theories and Supergravity", Adv. Theor. Math. Phys. {\bf 2} (1998) 231;
S.S.~ Gubser, I.~R.~ Klebanov, A.~M.~ Polyakov, "Gauge Theory
Correlators from Non-Critical String Theory", Phys. Lett. {\bf B428}
(1998) 105;
E.~Witten, "Anti De Sitter Space And Holography",
Adv. Theor. Math. Phys. {\bf 2} (1998) 253.}
\lref\Townsend{P.K.~Townsend, ``PhreMology--calibrating M-branes'',
hep-th/9911154, contribution to {\it Strings '99} proceedings.}
\lref\BST{H.J.~Boonstra, K.~Skenderis, P.K.~Townsend,
``The domain-wall/QFT correspondence'', JHEP {\bf 9901} (1999) 003.}
\lref\BBHS{K.~Behrndt, E.~Bergshoeff, R.~Halbersma and J.P.~van der Schaar,
``On domain-wall/QFT dualities in various dimensions'', Class.Quant.Grav.
{\bf 16} (1999) 3317, hep-th/9907006. }
\lref\DOW{R.~Donagi, B.A.~Ovrut, D.~Waldram, ``Moduli Spaces of Fivebranes
on Elliptic Calabi-Yau Threefolds'', JHEP {\bf 9911} (1999) 030.}
\lref\Michelson{J.~Michelson, ``Compactifications of Type IIB
Strings to Four Dimensions with Non-trivial Classical Potential",
Nucl. Phys. {\bf B495} (1997) 127.}
\lref\Anselmi{D.~Anselmi, ``Exact results on quantum field theories
interpolating between pairs of conformal field theories", hep-th/9910255.}
\lref\Hernandez{R.~Hernandez, ``Calibrated Geometries and
Non Perturbative Superpotentials in M-Theory", hep-th/9912022.}
\lref\tHooft{G.~t'Hooft, "A Planar Diagram Theory for Strong Interactions",
Nucl. Phys. {\bf B72} (1974) 461.}
\lref\Polyakov{A.M.~Polyakov, "String Theory and Quark Confinement",
Nucl. Phys. Proc. Suppl. {\bf B68} (1998) 1.}
\lref\PS{J. Polchinski and A. Strominger, `New Vacua for
Type II String Theory,'' Phys. Lett. {\bf B388} (1996) 736.}
\lref\BB{K.~Becker, M.~Becker, ``M-Theory on Eight-Manifolds",
Nucl.Phys. {\bf B477} (1996) 155.}
\lref\CCAF{A.~C.~Cadavid, A.~Ceresole, R.~D'Auria, and S.~Ferrara,
Phys. Lett. {\bf B357} (1995) 76.}
\lref\MQCD{E.~Witten, ``Branes And The Dynamics Of QCD",
Nucl.Phys. {\bf B507} (1997) 658.}
\lref\LOSW{A.~Lukas, B.A.~Ovrut, K.S.~Stelle, D.~Waldram,
``The Universe as a Domain Wall", Phys.Rev. {\bf D59} (1999) 086001;
``Heterotic M-theory in Five Dimensions", Nucl.Phys. {\bf B552} (1999) 246.}
\lref\Stelle{K.S.~Stelle, ``Domain Walls and the Universe", hep-th/9812086.}
\lref\RS{L.~Randall and R.~Sundrum,
``An Alternative to Compactification", hep-th/9906064;
L.~Randall and R.~Sundrum,
``A Large Mass Hierarchy from a Small Extra Dimension", hep-ph/9905221.}
\lref\GPPZ{L.~Girardello, M.~Petrini, M.~Porrati and A.~Zaffaroni,
``Novel local CFT and exact results on perturbations of N = 4
super Yang-Mills from AdS dynamics'', JHEP {\bf 12}
(1998) 022, hep-th/9810126 .}
\lref\KPW{A.~Khavaev, K.~Pilch and N.P.~Warner,
``New vacua of gauged N = 8 supergravity in five dimensions'',
hep-th/9812035.}
\lref\FGPW{D.Z.~Freedman, S.S.~Gubser, K.~Pilch and N.P.~Warner,
``Renormalization group flows from holography supersymmetry
and a  c-theorem'', hep-th/9904017.}
\lref\CGR{M.~Cvetic, S.~Griffies and Soo-Jong Rey,
``Static domain walls in N=1 supergravity'', Nucl.Phys. {\bf B381}
(1992) 301, hep-th/9201007.
M.~Cvetic and H.H.~Soleng, ``Supergravity domain walls'',
Phys.Rept. {\bf 282} (1997) 159, hep-th/9604090.}
\lref\BC{K.~Behrndt, M.~Cvetic, ``Supersymmetric Domain-Wall World from
D=5 Simple Gauged Supergravity'', hep-th/9909058.}
\lref\Behrndt{K.~Behrndt, ``Domain walls of D=5 supergravity and
fixpoints of N=1 Super Yang Mills", hep-th/9907070.}
\lref\GST{M.~Gunaydin, G.~Sierra, P.K.~Townsend,
``The N=2 Maxwell-Einstein Supergravity Theories:
Their Compact and Noncompact Gaugings and Jordan Algebras",
Proc. of Nuffield Workshop (1985) 0367.}
\lref\Wflux{E. Witten, ``On Flux Quantization In M-Theory And
The Effective Action", J. Geom. Phys. {\bf 22} (1997) 1.}
\lref\GVW{S.~Gukov, C.~Vafa and E.~Witten, ``CFT's From
Calabi-Yau Four-folds", hep-th/9906070.}
\lref\Gukov{S.~Gukov, ``Solitons, Superpotentials and Calibrations",
hep-th/9911011.}
\lref\SVW{S. Sethi, C. Vafa, and E. Witten,
``Constraints on Low-Dimensional String Compacti\-fi\-cations",
Nucl. Phys. {\bf B480} (1996) 213.}
\lref\Lerche{W.~Lerche, ``Fayet-Iliopoulos Potentials from
Four-Folds", JHEP {\bf 9711} (1997) 004.}
\lref\DRS{K.~Dasgupta, G.~Rajesh, S.~Sethi,
``M Theory, Orientifolds and G-Flux", hep-th/9908088.}
\lref\HL{R.~Harvey and H.B.~Lawson, Jr., ``Calibrated geometries",
Acta Math. {\bf 148} (1982) 47.}
\lref\SVV{M.~Shifman, A.~Vainshtein, M.~Voloshin, ``Anomaly and
quantum corrections to solitons in two-dimensional theories with
minimal supersymmetry", Phys. Rev. {\bf D59} (1999) 045016.}
\lref\Strominger{A.~Strominger, ``Loop Corrections to the Universal
Hypermultiplet", Phys. Lett. {\bf B421} (1998) 139.}
\lref\KN{I.R.~Klebanov, N.~A.~Nekrasov, ``Gravity Duals of
Fractional Branes and Logarithmic RG Flow", hep-th/9911096.}
\lref\LNV{A.~Lawrence, N.~Nekrasov, C.~Vafa, ``On Conformal
Theories in Four Dimensions", Nucl. Phys. {\bf B533} (1998) 199.}
\lref\TV{T.R.~Taylor, C.~Vafa, ``RR Flux on Calabi-Yau and Partial
Supersymmetry Breaking", hep-th/9912152.}
\lref\HP{M.~Wijnholt, S.~Zhukov, ``On the Uniqueness of black hole
 attractors'', hep-th/9912002.}
\lref\FK{S.~Ferarra, R.~Kallosh, ``Supersymmetry and attractors''
Phys.Rev. {\bf D54} (1996)1514,  hep-th/9602136 ,
``Universality of supersymmetric attractors'', Phys.Rev. {\bf D54}
(1996) 1525, hep-th/9603090.}
\lref\MOR{D.~Morrison, ``Beyond the Kaehler cone'', alg-geom/9407007.}
\lref\GPPZZ{L.~Giradello, M.~Petrini, M.~Porrati, A.~Zaffaroni,
``The supergravity dual of N=1 super Yang-Mills theory'',
hep-th/9909047.}
\lref\KA{R.\ Kallosh, A.\ Linde, M.\ Shmakova,
``Supersymmetric multiple basin attractors'',
JHEP {\bf 9911} (1999) 010 , hep-th/9910021}
\lref\CLP{M.\ Cvetic, H.\ L\"u, C.N.\ Pope, ``Domain walls and massive
gauged supergravity potentials'', hep-th/0001002}
\lref\Duff{M.J.\ Duff, ``TASI Lecture on branes, black holes and
anti-de Sitter space'', hep-th/9912164.}
\lref\Ferrara{A.C.~Cadavid, A.~Ceresole, R.~D'Auria, S.~Ferrara,
``11-Dimensional Supergravity Compactified on Calabi-Yau
Threefolds'', Phys.Lett. {\bf B357} (1995) 76.}
\lref\Kallosh{R.~Kallosh, A.~Linde, ``Supersymmetry and
the Brane World'', hep-th/0001071.}

%%%%%%%%%%%%%%%%%%%%%%%%%%%%%%%%%%%%%%%%%%%%%%%%%%%%%%%%%%%%%%%%%%%%%%

\Title{\vbox{\baselineskip12pt
\hbox{hep-th/0001082}
\hbox{PUPT-1909}
\hbox{ITEP-TH-83/99}
\hbox{CALT-68-2252}
\hbox{CITUSC/00-001}}}
{{\vbox{\centerline{Domain Walls and Superpotentials from}
\medskip
\centerline{M Theory on Calabi-Yau Three-Folds}}}}
\centerline{Klaus Behrndt \quad and \quad Sergei Gukov}
\bigskip
\bigskip
\centerline{\it behrndt@theory.caltech.edu \ , \ gukov@theory.caltech.edu}
\bigskip
\centerline{\it Department of Physics}
\centerline{\it California Institute of Technology}
\centerline{\it Pasadena, CA 91125, USA}
\bigskip
\centerline{\it CIT-USC Center For Theoretical Physics}
\centerline{\it University of Southern California}
\centerline{\it Los Angeles, CA 90089-2536, USA}
\vskip .3in
\centerline{\bf Abstract}

Compactification of M theory in the presence of $G$-fluxes
yields $\CN=2$ five-dimensional gauged supergravity
with a potential that lifts all supersymmetric vacua.
We derive the effective superpotential directly from
the Kaluza-Klein reduction of the eleven-dimensional action
on a Calabi-Yau three-fold and compare it with the superpotential
obtained by means of calibrations.
We discuss an explicit domain wall solution, which
represents five-branes wrapped over holomorphic cycles.
This solution has a ``running volume'' and we comment on
the possibility that quantum corrections provide a lower bound
allowing for an $AdS_5$ vacuum of the 5-dimensional supergravity.

\Date{}

%%%%%%%%%%%%%%%%%%%%%%%%%%%%%%%%%%%%%%%%%%%%%%

\newsec{Introduction}

In this paper we study compactification of M theory on Calabi-Yau
three-folds in the presence of background $G$-fluxes.
If there were no $G$-fluxes, the effective field theory
would be $\CN=2$ five-dimensional supergravity interacting
with some number of hypermultiplets and vector multiplets
whose scalar fields parametrize a manifold $\CM$.
Turning on a non-trivial $G$-flux generates effective superpotential
in the five-dimensional theory, which is related to
gauging of global isometries of the scalar manifold $\CM$.
If the potential in the five-dimensional theory allows for
isolated extrema, the vacuum is given by a space-time of constant
negative curvature (i.e.\ an AdS space) and such a theory
is relevant to the AdS/CFT correspondence \ADS. 
On the other hand, we find that the potential is a monotonic
function of volume scalar, and this ``run-away'' case is
relevant to the generalization of the AdS/CFT correspondence,
the so-called domain wall/QFT correspondence \refs{\BST,\BBHS}.
However, if the Calabi-Yau space has positive Euler number
then the running of the volume is bounded by quantum
corrections \Strominger, so that five-dimensional supergravity
has an AdS$_5$ vacuum.
Below we list various applications which motivated our work.

Over the past year, domain walls as solutions of 5-dimensional gauged
supergravity that interpolate between different vacua has been a
subject of intensive research; for earlier work on domain wall
solution of 4-dimensional supergravity see \CGR.  Most of them are
dealing with the maximal supersymmetric case like in
\refs{\GPPZ,\FGPW} and many subsequent papers, but also the least
supersymmetric case has been discussed
\refs{\LOSW,\Stelle,\Behrndt,\BC}.  For a recent review see \Duff\ and
a discussion that of a running breathing mode
is given in \refs{\CLP}.

A model of domain wall universe was used by Randall
and Sundrum \RS\ to address the hierarchy problem in a novel way,
alternative to compactification.  An interesting feature of
Randall-Sundrum construction is that gravity is localized on the
domain wall (or D-brane) by a suitable gravitational potential.  The
original construction of \RS\ is purely classical and is based on a
non-supersymmetric example of a domain wall which interpolates between
two regions of five-dimensional space-time with negative cosmological
constant.  However, locally AdS form of the vacua on each side of the
wall suggests that there must be a corresponding supersymmetric
solution.  For a related recent work see \Kallosh. 
Furthermore, motivated by the celebrated D-brane
construction of MQCD \MQCD, one would like to embed a model a la
Randall-Sundrum in string theory or M theory to learn about
non-perturbative effects in the theory on the domain wall.
Even though we will not be able to solve this problem in the full
generality, we hope that our study of domain walls constructed from
M5-branes wrapped on holomorphic curves inside a Calabi-Yau space
will be a useful step in this direction. 
This configuration has been discussed in the
heterotic M-theory compactification  in \LOSW. 

Another line of research which motivated this paper is a quest
for new supersymmetric vacua in compactifications of string theory
or M theory on Calabi-Yau manifolds with background fluxes.
Since the flux has to be quantized, its different values correspond to
distinct disconnected components in the space of supersymmetric vacua.
Therefore, if we call $\CF$ the background flux and $\CM_{\CF}$
the corresponding component of the moduli space, the total
space of vacua looks like:
\eqn\mdisc{ \CM = \coprod_{\CF} \CM_{\CF} }
The component $\CM_0$ is equivalent, at least locally,
to the moduli space, $\CM(Y)$, of the Calabi-Yau space $Y$.
The other components are isomorphic to some subspaces
in the Calabi-Yau moduli space, $\CM_{\CF \ne 0} \subseteq \CM(Y)$,
such that all points in $\CM_{\CF}(Y)$ correspond
to the values of Calabi-Yau moduli which lead to supersymmetric
compactifications on $Y$ with a given flux $\CF$.
For example, when $Y$ is a Calabi-Yau three-fold
new supersymmetric vacua can be found at some special
(conifold) points of the moduli space \refs{\PS,\Michelson}.
For a Calabi-Yau four-fold there is usually more possibility
to turn on background fluxes which do not break supersymmetry
further \refs{\BB,\GVW,\DRS,\Gukov}.
Since the value of the flux $\CF$ jumps across a brane
of the appropriate dimension, this brane wrapped over a
supersymmetric cycle in $Y$ can be identified with a BPS
domain wall interpolating between different components in \mdisc.
This interpretation was used in \refs{\GVW,\Gukov} to deduce
the effective superpotential $W(\CF)$ generated by a flux $\CF$
in compactification on a Calabi-Yau four-fold $Y$, such that
its minima over $\CM(Y)$ reproduce the space of vacua $\CM_{\CF}$.
Using a more general argument which also applies to
compactifications on $G_2$ and $Spin(7)$ manifolds, one finds
the following universal formula for the effective superpotential
in terms of calibrations of % the space-time
$Y$ \Gukov:
\eqn\calibr{W = \sum \int_{Y} ({\rm fluxes}) \wedge ({\rm calibrations})}

The paper is organized as follows.
In the next section we perform a Kaluza-Klein reduction
of the eleven-dimensional supergravity action
on a Calabi-Yau three-fold with a $G$-flux.
Among other things we find that all supersymmetric vacua
of the five-dimensional theory are lifted by the effective
superpotential which does not have stable minima.
In section 3 we rederive the same result identifying BPS
domain walls with five-branes wrapped over holomorphic curves,
and argue that the formula \calibr\ can be also applied
to compactifications on Calabi-Yau three-folds.
In section 4 we explicitly construct domain wall solutions
in the effective $D=5$ $\CN=2$ gauged supergravity which correspond
to M5-branes wrapped over holomorphic curves in the Calabi-Yau space.
The discussion in section 2 and 4 is in part parallel to the work \LOSW,
which we extend by the inclusion of quantum corrections 
yielding an AdS vacuum solution.

%%%%%%%%%%%%%%%%%%%%%%%%%%%%%%%%%%%%%%%%%%%%%%%%%%%%%%%%%%%%%%%%

\newsec{Compactification of M Theory on Calabi-Yau
Three-Folds with $G$-Fluxes}

In this section we  perform the compactification of M theory
on a Calabi-Yau three-fold $Y$ with a $G$-flux.  For the Kaluza-Klein 
reduction of the eleven-dimensional action:
\eqn\maction{ S_{11} = {1 \over 2} \int d^{11}x \sqrt{-g} R -
{1 \over 2} \int \Big[ {1 \over 2} G \wedge * G +
{1 \over 6} C \wedge G \wedge G \Big] }
we follow the standard procedure, which lead to gauged $\CN=2$
supergravity in five dimensions as discussed in \refs{\LOSW,\Stelle}. 
The latter theory has a potential for the scalar fields $X^I$
that play the role of local coordinates on the moduli space
of K\"ahler deformations of $Y$.
Unfortunately, the scalar potential always exhibits a run-away behavior,
so that compactification of M theory on $Y$ with non-zero $G$-flux
does not lead to new vacua in the effective five-dimensional theory.
It is worth mentioning that most of the material in this section
is not new and has appeared in the literature in various form.
In particular, we follow the steps of \Ferrara\ where analogous
compactification on Calabi-Yau three-folds without $G$-fluxes
was studied. In the context of Type II string theory compactifications
with background fluxes were discussed in the work \PS\ where
similar results were found. Closer to the subject of our paper is
the work \LOSW\ where compactification of M theory with a $G$-flux
was investigated and the induced superpotential was derived.
In order to make the paper self-consistent, below we perform
once again all the steps of the Kaluza-Klein reduction
in the form that will be convenient later.

The Kaluza-Klein reduction on a Calabi-Yau three-fold $Y$ yields
$h^{1,1}$ abelian gauge fields entering $h^{1,1}-1$ vector multiplets
and a gravity multiplet \Ferrara. The vector fields come from the light
modes of the 3-form field $C$ in eleven dimensions.
Namely, for the field strengths we have a decomposition:
\eqn\gone{ G \sim dA^I \wedge \omega_I}
where $\omega_I \in H^{(1,1)}(Y)$ is a basis of $(1,1)$-forms.  Each
vector multiplet contains besides the gaugino a real scalar which
comes from the reduction of the internal metric $g_{a \bar b} = -i \,
t^I (\omega_I)_{a \bar b}$, where $t^I$ are the K\"ahler moduli.
Identifying expectation values of the vector multiplet scalar
fields with $t^I$, we can write the K\"ahler form as follows:
\eqn\kform{ \CK = t^I \omega_I}

As we will see below, the scalar parameterizing the volume of
the Calabi-Yau decouples from the vector multiplets and enters
the universal hypermultiplet. This volume scalar is defined by:
$$
{\cal V} = \int \sqrt{g_{Y}} =
{1 \over 3!} \int \CK \wedge \CK \wedge \CK =
{1 \over 6} C_{IJK} t^I t^J t^K
$$
and the scalars $\phi^A$ ($A=1\ldots h^{1,1}-1$)
entering the vector multiplets are obtained from
\eqn\mfold{1 = {1 \over 6} C_{IJK} X^I X^J X^K \qquad {\rm with}
\qquad t^I = {\cal V}^{1\over 3} X^I }
i.e.\ $X^I = X^I(\phi^A)$.
In what follows we denote $\CM$ the manifold parameterized
by the scalar fields $\phi^A$, see figure below.

In addition to the volume scalar the universal hypermultiplet contains a
real scalar which is dual to the 4-form field in 5 dimensions:
\eqn\gtwo{ G \sim dC_3 \sim \ ^{\star}da }
and a complex scalar coming from:
\eqn\gthree{G \sim dm \wedge \Omega + d\bar m \wedge \bar \Omega}
In addition to these scalars further scalars are related to non-trivial
elements of $H^{(2,1)} (Y)$, which build up the remaining hyper
multiplets.  These fields are not important for our analysis, so, we
will ignore them.

\ifig\pic{Scalar components $\phi^A$ of vector multiplets
parametrize the space $\CM$ defined by the hypersurface equation \mfold.
At extrema of $W$, the normal vector $X_I$ has to be parallel to the
flux vector $\alpha_I$.}
{\epsfxsize2.5in\epsfbox{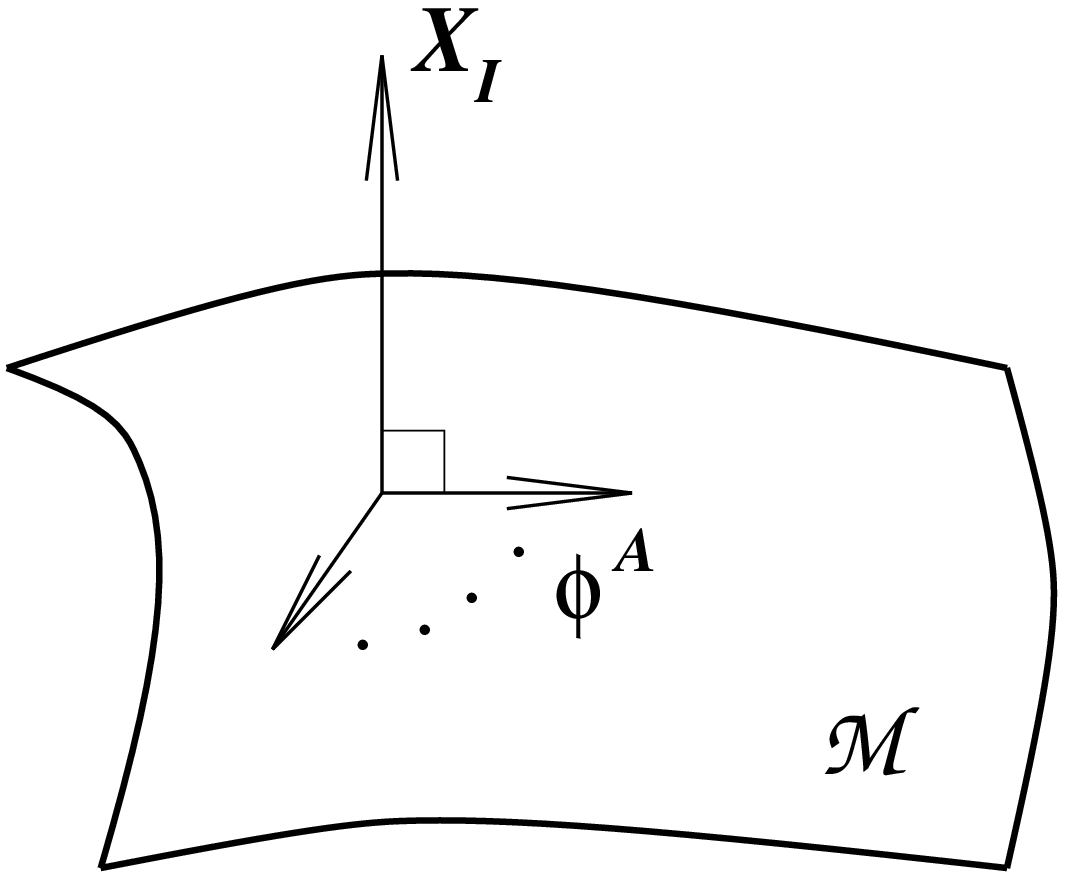}}

In order to obtain the canonical Einstein-Hlibert term in 5d, we have to
perform a Weyl rescaling (Einstein versus string frame) which is related
to the volume of the internal space. In five dimensions this rescaling
is given by:
$$
ds^2_E = {\cal V}^{2\over 3}  ds^2_{str} \qquad , \qquad \sqrt{g_{str}}
= \sqrt{g_E} \; {\cal V}^{- 5\over 3}.
$$
with the eleven-dimensional metric written as $ds_{11}^2 = ds_{str}^2 +
ds^2_{CY}$.
% which is a direct product
% of the internal space metric and the metric of the non-compact
% five-dimensional space-time.
Combining this rescaling with the rescaling of the scalars
in \mfold, the reduction of the Ricci scalar yields \Ferrara:
$$
S_5 = \int \Big[ {1 \over 2} R - {1 \over 2} G_{IJ}(X)\,
\partial X^I \partial X^J
 - {1 \over 2} {\partial {\cal V} \partial {\cal V} \over {\cal V}^2 } \Big]
$$
where one has to use the relation $G_{IJ}(X) \, X^I \partial X^J = 0$
and $G_{IJ}$ as function of the $t^I$ coordinates is defined by:
$$
G_{IJ}(t) = {i \over 2 {\cal V}} \int \omega_I \wedge \ ^{\star} \omega_J
= - {1 \over 2}  \Big[ {C_{IJK} t^K \over {\cal V}} - {1 \over 4}
{(C_{IKL} t^K t^L)(C_{JMN} t^M t^N) \over {\cal V}^2} \Big]
$$
After rescaling into $X$ coordinates it takes the form:
\eqn\gg{G_{IJ}(t) = {\cal V}^{-{2 \over 3}} G_{IJ}(X)}
with
$$
G_{IJ}(X) = - {1 \over 2}  \Big[ C_{IJK} X^K  - {1 \over 4}
{(C_{IKL} X^K X^L)(C_{JMN} X^M X^N) } \Big] \ .
$$
Notice, that ${3\over 2} G_{IJ} X^J = X_I$ is the normal vector and
$\partial_A X^I$ are tangent vectors on the scalar manifold $\CM$,
as shown on Fig.1.

In order to perform the reduction of the $G\wedge ^{\star} G$ term in \maction,
consider the gauge field term \gone\ which in five
dimensions becomes:
$$
\int \sqrt{g_{str}} \, {\cal V} \, G_{IJ}(t) F_{\mu\nu}^I F^{J}_{\mu' \nu' }
g_{str}^{\nu\nu'} g_{str}^{\mu\mu'}
= \int \sqrt{g_E} \, G_{IJ}(X) F_{\mu\nu}^I F^{J}_{\mu' \nu' }
g_{E}^{\nu\nu'} g_{E}^{\mu\mu'}
$$
and \gtwo\ yields:
$$
\int \sqrt{g_{str}} \, {\cal V} \, (dC_3)^2 = \int \, \sqrt{g_{E}}
\, {\cal V}^{2} \, (dC_3)^2 \ .
$$

We are looking for potentials that we can obtain from
non-trivial $G$-fluxes.
The flux quantization condition can be written in the following
form\foot{In general, the periods $\alpha_I$ are only required
to be half-integer \Wflux.}:
\eqn\qcond{\int_{Y} G_{flux} \wedge \omega_I = \alpha_I = {\rm integer}}
Since the internal space remains a Calabi-Yau, in particular
Ricci-flat,  the only source for a potential comes from the $G^2$ term.
The topological term contains a derivative in the uncompactified space
and therefore cannot give a potential.
Let us consider the example discussed in \refs{\LOSW,\Stelle}:
\eqn\oflux{ G_{flux} = {1 \over \CV} \alpha^I \ ^{\star}\omega_I }
with $\alpha^I = G^{IJ}(t) \, \alpha_J$ in agreement with \qcond. This yields:
$$
\int_{M_{11}} G_{flux} \wedge \ ^{\star} G_{flux} =
2 \int_{M_5} \sqrt{g_{str}}
\, {1 \over  {\cal V}} \, \alpha^I \alpha^J G_{IJ}(t) = 2 \int_{M_5} \sqrt{g_E}
\, {1\over {\cal V}^2} \, \alpha_I \alpha_J G^{IJ}(X)
$$

Note the difference between $G_{IJ}(t)$ and $G_{IJ}(X)$, {\it cf.} \gg.
Writing the volume scalar as:
$$
{\cal V} = e^{-2 \varphi}
$$
the potential becomes:
\eqn\potential{
V(X, \varphi) = e^{4 \varphi} \Big(\alpha_I \alpha_J G^{IJ}(X) \Big)}
A potential of this form was originally found in \LOSW.

If we include the $G$-flux in the topological term we obtain
after compactification
\eqn\topolog{
\int_{M_{11}} G \wedge C \wedge G = \int_{M_5} G \wedge A^I
\int_{Y}\omega_I \wedge ^{\star} \omega_J \alpha^J
= \int_{M_5} G \wedge A^I \alpha_I }
and after dualization the 4-form $G$, the corresponding scalar $a$
becomes charged under the gauge field $A^I \alpha_I$.
This effect, as well as the generation of the potential \potential\
can also be understood from the reduction of the eleven-dimensional
supersymmetry transformations:
\eqn\msusy{\delta e^A_M = i \overline  \eta \Gamma^A \psi_M, \quad
\delta C_{MNP} = 3i \overline \eta \Gamma_{[MN} \psi_{P]}.}
\eqn\susy{ \delta \psi_M = \nabla_M \eta - {1 \over 288}
({\Gamma_M}^{PQRS} - 8 \delta_M^P \Gamma^{QRS}) G_{PQRS} \eta.}
Here the supersymmetry parameter $\eta$ is an eleven-dimensional
Majorana spinor. Under the split 11=5+6, we decompose it as
$\eta = \epsilon \otimes \xi$ where $\epsilon$ is
an anti-commuting spinor in five non-compact dimensions.

If there were no background $G$-fluxes, then the resulting
supersymmetry transformations in five dimensions would correspond to
the usual (not gauged) supergravity theory which does not allow a
scalar potential.  This supergravity would have $SU(2)$ $R$-symmetry
group.  Gauging a $U(1)$ subgroup of the $R$-symmetry group, one
obtains a gauged supergravity first found by Gunaydin, Sierra and
Townsend \GST.
% This latter theory does have a scalar potential of the form \potential.
They considered only vector multiplets which effectively means that
the volume of the internal space is assumed to be fixed. In this
case  the supersymmetry transformations in the gauged
theory differ only by the extra terms:
\eqn\dgaugino{ \delta \lambda^{i A} = P^A \delta^{ij} \epsilon_j}
in the gaugino variation, and:
\eqn\dgravitino{ \delta \psi_{\mu}^i =
{i \over 2 \sqrt{6}} P_0 \gamma_{\mu} \delta^{ij} \epsilon_j }
in the variation of gravitino. In our notation $P^A \sim \partial^A
W$ and $P_0 \sim W$. On the other hand, allowing for
general $G$-fluxes also yields a dynamical volume scalar which is
equivalent to a rescaling of $W$ combined with an additional term
in the potential $V$.

It is easy to see, for example, how \dgravitino\ comes from
the supersymmetry transformations \susy\ with a $G$-flux.
If the background field $G$ has non-zero components only in
the internal space, then only the first term in brackets is
relevant\foot{In order to obtain \dgravitino, we also make
a decomposition of gamma-matrices
$\Gamma_{\mu} = \gamma_{\mu} \otimes \gamma_7$ and
$\Gamma_m = 1 \otimes \gamma_m$ in the formula \susy.
The eleven-dimensional gamma-matrices $\Gamma^M$ are hermitian
for $M=1,\ldots, 10$ and anti-hermitian for $M=0$.}.
The extra term \dgaugino\ can be obtained in a similar way.

To summarize, in the large volume limit the effective
five-dimensional gauged supergravity action reads \LOSW:
\eqn\daa{\eqalign{ S  \sim & \int \sqrt{G}
\Big[ {1 \over 2} R - g^2V -  {1 \over 4}
G_{IJ} F_{\mu\nu}^I F^{\mu\nu J} -
{1 \over 2} g_{AB} \partial_{\mu} \phi^A \partial^\mu \phi^B  -
{1 \over 2} h_{rs} D_{\mu} q^r D^{\mu} q^s \Big] \cr & +
\int C_{IJK} F^I \wedge F^J \wedge A^K }}
with
\eqn\brd{ g_{AB} = \partial_A X^I \partial_B X^J G_{IJ} }
subject to the constraint ${1 \over 6} C_{IJK} X^I X^K X^J = 1$.
Using the convention of \LOSW\ 
the metric  of the universal hypermultiplet $h_{rs}$ is given
by 
%\foot{Here we use the conventions of \LOSW.}
\eqn\dab{\eqalign{
h_{rs} dq^r dq^s = {1 \over 4 {\cal V}^2} d{\cal V}^2 + {1 \over
2 {\cal V}^2} \Big[ da + i (m \, d\bar m - \bar m \, d m)
\Big]^2 + {1 \over \CV} \, dm d \bar m = u \bar u + v \bar v
}}
where
\eqn\daaab{
u = {d m  \over \sqrt{\cal V}} \qquad , \qquad
v = {1 \over 2 {\cal V}} (d{\cal V} + i da + m d\bar m - \bar m dm)
}
and this metric parameterizes the coset $SU(2,1)/U(2)$. Recall, {$\cal V$} is
the volume scalar, the axionic scalar $a$ comes from the dualization
of the five-dimensional 3-form field and the complex scalar $m$
was introduced in \gthree. Notice, due to the non-trivial flux only
the axionic scalar $a$ becomes charged:
\eqn\dac{
D_{\mu} q^r = \big\{ \partial_{\mu} {\CV} \ , \
\partial_{\mu} a + A^I_{\mu} \alpha_I \ , \ \partial_{\mu} m \ , \
\partial_{\mu} \bar m \big\}
}
In the supergravity theory this corresponds to a gauging
of the axionic shift symmetry $a \rightarrow a + const$.
In order to understand the structure of the potential,
we have to understand the gauging on the supergravity side
\LOSW. Obviously, the $G$-fluxes correspond to a gauging
along the Killing vector:
\eqn\dad{ k = \partial_a = {i \over 2 {\cal V}}
(\partial_v - \partial_{\bar v}) }
The Killing prepotentials have the following form:
\eqn\dae{
 {\cal P}_I =\pmatrix{ -{i \over 4 {\cal V} } \alpha_I & 0 \cr
                      0 & {i \over 4 {\cal V}} \alpha_I }
}
and obey the relations:
\eqn\dag{
k^u_I {\cal K}_{uv} = \nabla_s {\cal P}_I = \partial_v {\cal P}_I +
[\omega_v , {\cal P}_I] }
where $\omega_v$ is the $v$ component of the
SU(2) connection and ${\cal K}_{uv}$ is the triplet of K\"ahler
forms\foot{Here $J^x$ denotes the triplet of complex structures.}
${\cal K}^x_{uv} = h_{uw} (J^x)_v^{\;w}$:
\eqn\daf{
\omega = \pmatrix{ {1 \over 4} (v - \bar v) & -u \cr
u & -{1 \over 4}(v - \bar v) }
 \ , \ {\cal K} = \pmatrix{{1 \over 2}(u \wedge \bar u - v \wedge \bar v)
& u \wedge \bar v \cr v \wedge \bar u & - {1 \over 2}( u \wedge \bar u
- v \wedge \bar v)} }

The gauging fixes the potential to be of the following form:
\eqn\daf{\eqalign{
V & =  4 tr({\cal P}_I {\cal P}_J)\Big[ 2 X^I X^J - G^{IJ} \Big]
+ 2 X^I X^J h_{uv} k_I^u k_J^v \cr
&= 4 e^{4 \varphi} \, \Big[g^{AB} \partial_A W \partial_B W
- {4 \over 3} W^2 \Big] + 2 e^{4 \varphi} \, W^2 |k|^2 }}
where
\eqn\dbf{
W \equiv {\alpha_I X^I}
}
Inserting this expression into \daf, one finds that the last two terms
in the first line cancel and the first term agrees with the
potential in \potential.

So far our discussion was purely classical. However, it is
very easy to incorporate corrections due to the non-minimal
terms in the action \maction. It was found by Strominger
\Strominger\ that corrections due to the terms proportional
to the fourth power of the Riemann curvature simply lead
to the shift (redefinition of the dilaton field):
\eqn\uhypershift{ e^{- 2 \varphi} \to e^{- 2 \varphi} +
{\chi (Y) \over 15 \cdot 2^{10} \cdot \pi^8} }
where $\chi (Y)$ is the Euler number of the Calabi-Yau space $Y$. In
string theory this would be a one-loop correction to the metric on the
moduli space of the universal hypermultiplet. We note that as long as we
consider only the universal hypermultiplet, we do not expect further
corrections especially no instanton corrections. In addition, the shift
\uhypershift\ effectively puts a lower bound on the ``quantum'' volume of
the Calabi-Yau space, which leads to some qualitative changes of the
supergravity solutions. As we will see in the section 4, the domain
wall describes a supergravity solution with monotonically decreasing
volume and if it eventually reaches this lower bound, we can keep the
volume constant and allowing afterwards only internal deformations as
described by the scalars in the vector multiplets. As consequence, at
the point where this ``quantum'' volume is reached, the volume scalar
effectively decouples from our supergravity solution and the scalars
in the vector multiplets settle down at the extremum of the
potential. This configuration is described by an AdS vacuum.
Of course, this interpretation makes sense only if $\chi (Y) >0$.

%%%%%%%%%%%%%%%%%%%%%%%%%%%%%%%%%%%%%%%%%%%%%%%%%%%%%%%%%%%%%%%%%%

\newsec{More Superpotentials From Calabi-Yau Calibrations}

In this section we discuss a way to derive the effective
superpotentials via identification of BPS domain walls
with branes wrapped over supersymmetric cycles.
Although in this paper we are mainly interested in
M theory compactifications on Calabi-Yau three-folds,
we will also consider string theory compactifications.

Let us start with a general compactification of string theory
or M theory on a compact oriented manifold $Y$ to $(d+1)$
non-compact dimensions.
In other words, the (real) dimension of space $Y$ is equal
to $(9-d)$ in string theory, or $(10-d)$ in M theory.
Trying to keep the discussion as general as possible,
we make only a few minor assumptions about the geometry
of the space-time. Namely, we assume that compactification on $Y$
preserves some supersymmetry, so that it makes sense to talk
about BPS domain walls in $(d+1)$-dimensional effective theory.
Non-compact space-time is assumed to be a maximally symmetric
homogeneous space with zero or negative cosmological constant,
{\it i.e.} Anti de Sitter space or a Minkowski space.

Assuming further the existence of a $(d+k-1)$-brane in
the theory we start with, we can construct a BPS domain
wall in the effective field theory by wrapping this brane
over a supersymmetric $k$-cycle $\Sigma \subset Y$,
of course, if there is  one.
Indeed, a simple counting of dimensions shows that
the resulting object  should be codimension one in
the non-compact space-time.
Notice, supersymmetric branes that we consider represent
a magnetic source for some field strength in string theory
or M theory, depending on the model in question.
Let us call this field strength $\CF$.
Thus, as we move across the domain wall in $(d+1)$
dimensions, the field strength jumps, $\CF \to \CF + \Delta \CF$.
The change of the flux, $\Delta \CF$, is determined
by the geometry of the $(d+k-1)$-brane that we used
to construct the domain wall. Namely, we have:
\eqn\flux{\Delta \CF = \hat{[\Sigma]} }
where the cohomology class $\hat{[\Sigma]} \in H^* (Y,\Z)$
is Poincar\'e dual to the homology class $[\Sigma]$.

Let us now return to the BPS property of the domain wall.
Since BPS states have the least possible mass, and the domain
wall in question is represented by a $(d+k-1)$-brane wrapped
over $k$-dimensional cycle $\Sigma$, we conclude that $\Sigma$
should have the minimal volume in its homology class.
Due to this last property, calibrated geometries
introduced by Harvey and Lawson \HL\ turn out to be very
useful in a study of supersymmetric brane configurations
(see \Townsend\ for a review and a list of references).
Here we give only the definition of a calibrated
submanifold and refer the reader to the original
paper \HL\ for further details.
A closed $k$-form $\Psi$ is called a calibration if
its restriction to the tangent space $T_x \Sigma$
is not greater than the volume form of $\Sigma$ for
every submanifold $\Sigma \subset Y$.
By saying this we mean that
$\Psi \vert_{T_x \Sigma} \le {\rm vol} (T_x \Sigma)$
is satisfied provided that
$\Psi \vert_{T_x S} = c \cdot {\rm vol} (T_x S)$
for some real coefficient $c \le 1$.
Furthermore, if $\Psi \vert_{T_x \Sigma} = {\rm vol} (T_x \Sigma)$
for every point $x \in \Sigma$, the submanifold is
called a calibrated submanifold with respect to
the calibration $\Psi$. It follows that calibrated
submanifolds have the minimal volume in their homology class:
\eqn\svol{{\rm Vol}(\Sigma) = \int_{\Sigma} \Psi}

The last assumption we are going to make is that the mass
of our BPS domain wall in the $(d+1)$-dimensional effective theory
is determined by the usual BPS formula:
\eqn\bps{M_{BPS} = \vert \Delta W \vert}
where $W$ is the effective superpotential.
Then, combining the formulas \flux, \svol\ and \bps\ together
we obtain the following formula for the superpotential
generated by a flux $\CF \in H^*(Y)$:
\eqn\wone{W = \int_Y \Psi}

The approach via calibrated geometries that we have outlined above can
be applied to a computation of tree-level superpotentials induced by
background fluxes in compactifications of string theory and M theory
on Calabi-Yau manifolds \refs{\GVW, \Gukov, \TV}, and to the
derivation of membrane instanton superpotentials in M theory
compactifications on $G_2$ manifolds \Hernandez.
Although all the results agree with what one finds studying
the supersymmetry conditions, it would be also interesting
to derive the effective superpotentials directly from the
Kaluza-Klein reduction of the Lagrangian, {\it cf.} \TV.
It is clear that in the case of Calabi-Yau four-folds,
non-minimal terms like the anomaly term $C \wedge I_8(R)$ and,
perhaps, their supersymmetric completion must play an important role \SVW.

Compactifications on $Spin(7)$ manifolds
preserve only two real supercharges in the effective field theory.  It
was demonstrated in \SVV\ that the BPS mass condition \bps\ is modified in
such theories by the one-loop quantum anomaly $W \to W + {W'' \over 4
\pi}$.  Therefore, one might expect that the effective superpotential
induced by a four-form flux in compactification on $Spin(7)$ manifold
is given by the appropriate modifications of the formula \wone\ which
takes into account one-loop quantum anomaly. It would be interesting
to see this anomaly by a direct computation of the superpotential via
Kaluza-Klein reduction of the ten-dimensional supergravity action or
supersymmetry transformations.

In this paper we focus on the case where $Y$ is a Calabi-Yau
three-fold. There are two types of calibrations on Calabi-Yau
three-folds.  The first type of calibrations --- so-called K\"ahler
calibrations --- includes closed forms of even degree constructed from
various powers of the K\"ahler form $\CK$:
\eqn\ckahler{ \Psi = {1 \over p!} \CK^p }
Apart from the trivial examples corresponding to $p=0$ or
$p=3$, the submanifolds calibrated by such $\Psi$ are
holomorphic curves and divisors in $Y$.
The second type of calibrations --- the special Lagrangian calibration:
\eqn\cslag{ \Psi = {\rm Re} (\Omega) }
corresponds to special Lagrangian submanifolds in $Y$.
Here $\Omega \in H^{3,0} (Y)$ is the unique holomorphic 3-form.

Clearly, the formula \wone\ can be used in compactifications
of heterotic string theory on Calabi-Yau three-folds.
In this case, four-dimensional effective field theory has
$\CN=1$ supersymmetry. The only way to construct a BPS
domain wall in four non-compact dimensions is to consider
a five-brane wrapped around a special Lagrangian cycle in $Y$.
Since a five-brane is a source for the Neveu-Schwarz
three-form field strength, eq. \wone\ yields:
$$
W = \int_Y H \wedge \Omega
$$
We believe that this formula can be derived by the direct
arguments, similar to what we used in the previous section.

It turns out that the formula \wone\ can be also
applied to theories with larger supersymmetry.
Recently, Taylor and Vafa \TV\ studied the effect of
background fluxes in Type II string theory on
(non-compact) Calabi-Yau three-folds.
They found that it leads to partial supersymmetry
breaking via generation of the effective superpotential \wone\
and reconciled it with the results of \refs{\PS,\Michelson}.
In particular, in Type IIA string theory on a Calabi-Yau
three-fold $Y$ the effective superpotential induced
by the flux $\CF$ has the following form \TV:
$$
W = \int_Y e^{\CK} \wedge \CF
$$
which is exactly what follows from \wone\ with
the K\"ahler calibration \ckahler.

One goal of the present paper is to demonstrate that
effective superpotential of the form \wone\ is also
generated in compactification of M theory on
a Calabi-Yau space $Y$ with a four-form field flux $G$.
The resulting field theory in five dimensions has $\CN=2$ local supersymmetry.
In $\CN=2$ five-dimensional gauged supergravity theories all our assumptions,
including the BPS formula \bps, are justified by the relation between
central charge of $\CN=2$ supersymmetry algebra and the gravitino mass
as discussed in \refs{\CGR, \BC}.
Since the four-form field strength $G$ is the only possible flux
in M theory, the formula \wone\ predicts
the following simple superpotential
$W \sim \int_Y \CK \wedge G_{flux} = \alpha_I t^I$.
In the five-dimensional $\CN=2$ gauged supergravity theory
we expect the effective superpotential $W$ to be a function
of the scalar fields $X^I$ from vector multiplets, rather
than $t^I$ which also include a volume scalar $\CV$ from
the universal hypermultiplet. Since $\alpha_I$ are
integer numbers, after the appropriate rescaling we obtain
the following superpotential:
\eqn\wm{W = \alpha_I X^I}
which is nothing but the effective superpotential \dbf\
found in the previous section via direct Kaluza-Klein reduction.

Notice that variation of the potential \wm\ with respect
to the fields $X^I$ leads to the condition:
$$
G=0
$$
which means that there are no supersymmetric vacua in
compactification of M theory on Calabi-Yau three-folds
with non-trivial fluxes. In other words, the space of
supersymmetric vacua has only one component
corresponding to $\CF=0$, {\it cf.} \mdisc.

%%%%%%%%%%%%%%%%%%%%%%%%%%%%%%%%%%%%%%%%%%%%%%%%%%%%%%%%%%%%%%%%%%

\newsec{Domain Wall Solutions}

Motivated by \MQCD, one may hope to understand non-perturbative
effects in realistic models a la Randall-Sundrum via embedding
the corresponding domain wall solutions in M theory or string theory.
Since $\CN=2$ five-dimensional supergravity can be obtained from
compactification of M theory on a Calabi-Yau three-fold $Y$, it is
natural to assume that the domain wall is constructed out of M5-brane
wrapped over a holomorphic curve $\Sigma \subset Y$,  see also \LOSW.
Then, topology of $Y$ and $\Sigma$ determine the spectrum
of the low-energy theory on the five-brane, and the appropriate
embedding $\Sigma \to Y$ may give us a theory close to the Standard Model.
Note, because the curve $\Sigma$ is holomorphic in $Y$,
the effective four-dimensional theory has $\CN=1$ supersymmetry.

Interested in domain wall solutions in the five-dimensional
supergravity we write the metric as:
\eqn\dag{
ds^2 = e^{2U(y)} \Big[ -dt^2 +dx^2_1 + dx_2^2 +dx_3^2 \Big] + e^{-2\gamma
U(y)} dy^2 }
where the constant $\gamma$ fixes the coordinate system and will be
chosen later.  This ansatz contains no restrictions as long as we
regard the four-dimensional domain wall as a flat Minkowski space, but
this parameterization will enable us to find an analytic solution
below.  Keeping the flat Minkowski space means also that the solution
cannot carry electric and/or magnetic charges, but can carry a
topological charge given by the difference of the cosmological
constants. It is thus consistent to set all the gauge fields to zero.
Moreover, investigating the equations of motion coming from the
Lagrangian, we find that the complex scalar $m$ and the axion $a$
can be neglected because they do not show up in the potential. We will
keep all scalars $t^I$, i.e. the scalars of in the vector multiplets
$\phi^A$ and the volume scalar ${\cal V} = e^{-2\varphi}$.

%%%%%%%%%%%%%%%%%%%%%%%%%%%%%%%%%%%%%%%%%%%%%%%%%%%%%%%%%%

\subsec{Solution of the $5d$ Killing spinor equations}

%%%%%%%%%%%%%%%%%%%%%%%%%%%%%%%%%%%%%%%%%%%%%%%%%%%%%%%%%

To ensure supersymmetry we have to solve the Killing spinor
equations.  Since the gauge fields are trivial for our domain wall the
relevant variations are:
\eqn\daj{\eqalign{
\delta \psi_{\mu} = & \Big( \partial_{\mu} +
{1 \over 4} \omega_{\mu}^{ab} \Gamma_{ab}
+ {1 \over 2} g \, \Gamma_{\mu} \, e^{2\varphi} W \Big) \epsilon \ , \cr
\delta \lambda_A =  & \Big( - {i \over 2} g_{AB} \Gamma^{\mu}
\partial_{\mu} \phi^B + i\, {3 \over 2} \, g \, e^{2\varphi} \,
\partial_A W \Big) \epsilon \ ,\cr
\delta \zeta = & e^{2\varphi} \Big( - {i \over 2} \Gamma^{\mu} \partial_{\mu}
e^{-2 \varphi}  - i\, 3 \, g  \,X^I k_I \Big) \epsilon
}}
with $W= {\alpha_I X^I}$.
% $V_u^{\; a}$ is the vielbein for the metric $h_{uv}$.
The scalar fields $\phi^A$ parameterize the manifold $\CM$
defined by \mfold, and the only non-trivial hypermultiplet field
$e^{-2\varphi} = {\cal V}$ gives the Calabi-Yau volume.

Let us start with the gravitino variation $\delta \psi$.
For our ansatz of the metric, the only non-zero components
of the vielbeine and spin connection are:
\eqn\dak{
e^m = e^{U} dx^m \ , \ e^y = e^{-\gamma U} dy \qquad , \qquad
\omega^{my} = e^{(\gamma + 1)U} \, U' \, dx^m
}
where $m = 0,1,2,3$ and the corresponding gravitino variation
becomes:
\eqn\dal{
\delta \psi_m = \Big( {1 \over 2} e^{(\gamma + 1) U} U' \, \Gamma_m \Gamma_y
    + {1 \over 2} \Gamma_{m} g e^U  e^{2\varphi}\, W \Big) \epsilon
}
Using the projector $(1 + \Gamma_y) \epsilon = 0$ we find:
\eqn\dan{
g \, e^{2\varphi}\, W = e^{\gamma U} U'  \ .
}
{}From the $\delta \psi_y$ component:
\eqn\dbr{
0 = \delta \psi_y = \Big( \partial_y + {1 \over 2} e^{-\gamma U} \,
\Gamma_y \, g \, e^{2\varphi}\, W \Big) \epsilon
}
we obtain the Killing spinor after using \dan:
\eqn\dbq{
\epsilon = e^{U \over 2} \Big(1 - \Gamma_y \Big) \epsilon_0
}
where $\epsilon_0$ is any constant spinor.

Moreover, using \dan\  we can also solve the hyperino variation:
\eqn\dap{\eqalign{
0& =  \Big(- {1 \over 2} \Gamma^{\mu} \partial_{\mu} e^{-2\varphi(y)}
- 3 g W \Big) \epsilon \cr
& = \Big( {1 \over 2} e^{\gamma U} (e^{-2\varphi})' - 3 \,
e^{\gamma U} \, U' e^{-2\varphi} \Big)
 \epsilon \cr
& = {1\over 2} e^{\gamma U} e^{-2 \varphi} \;
 \Big( 2 \varphi' - 6 \, U'  \Big) \epsilon
}
}
and therefore
\eqn\daq{
e^{6U} =  e^{-2(\varphi - \varphi_0)} = \CV/\ell
}
where $\ell = e^{-2\varphi_0}$ is the integration constant. Finally, we come
to the gaugino variation $\delta \lambda_A$ which gives:
\eqn\dar{\eqalign{
0 & = -{i \over 2}  \Big( g_{AB} \Gamma^{\mu} \partial_{\mu}
\phi^B - 3 g \,  e^{2\varphi}\, \partial_A W \Big) \epsilon \cr
& = - {i \over 2} \Big( \Gamma^{\mu} \partial_A X^I \partial_B X^J
G_{IJ} \partial_{\mu} \phi^B - 3 g \, e^{2 \varphi} \,
\partial_A ({\alpha_I X^I }) \Big) \epsilon \cr
&= - {i \over 2} \partial_A X^I \; \Big( e^{\gamma U} {3 \over 2} \,
\partial_{y} X_I - 3 g \, e^{2\varphi} \, {\alpha_I} \Big)
\epsilon \ . }}
Because $\partial_A X^I$ defines tangent vectors,
the expression in brackets has to be proportional to the
normal vector $X_I$:
\eqn\das{
{3\over 2 }\, e^{\gamma U} (X_I)' - 3 g \, e^{2\varphi}\,
{\alpha_I } = - 3 e^{\gamma U} U' X_I = -{3 \over \gamma} (e^{\gamma U})'
X_I }
the coefficient on the rhs can be verified by contracting the equation
with $X^I$ and using \dan. Next, replacing $e^{2\varphi}$ by employing
\daq\ and taking $\gamma = - 4$ we get
\eqn\dat{
{1 \over 2} \partial_y \Big( e^{2U} X_I\Big) = g \alpha_I/ \ell
}
and thus the solution is
\eqn\dau{
X_I \equiv {1 \over 6} C_{IJK} X^J X^K = e^{-2U} {1 \over 3} \, H_I =
e^{-2U} { 1 \over 3 \ell} \Big( q_I + 6 g \alpha_I \, y \Big) }
where $q_I$ are arbitrary constants. This solution agrees with the one
derived in \LOSW, but notice also the close relationship to the
attractor equations \FK\ which extremize the supersymmetry central
charge, or in our case, the superpotential $W$ and which
state that at extrema of $W$ the normal vector $X_I$ becomes parallel
to the flux vector $\alpha_I$. These extrema are reached at
$y \rightarrow \pm \infty$ where the scalars $\phi^A$ becomes
constant and $W$ extremal, due to \daj. Remember, because of
the run-away behavior of the volume, extrema of $W$ are not
extrema of the supergravity potential $V$.

As we discussed at the end of section 2, quantum corrections yield a
lower bound for the volume, which is mainly given by the Euler number of
the Calabi-Yau space. So, if we assume that in this ``quantum'' region
the universal hypermultiplets effectively decouples and if we approximate
the volume by the lower bound $\CV = e^{-2\varphi_0} = \ell$, we find the
same solution for eq.\ \das, but with $\gamma = +2$. In this case the
spacetime metric becomes asymptotically anti de Sitter, which is expected
because for a fixed volume, the potential has extrema.

%%%%%%%%%%%%%%%%%%%%%%%%%%%%%%%%%%%%%%%%%%%%%%%%%%%%

\subsec{Domain walls from five-branes in $T^6$}

%%%%%%%%%%%%%%%%%%%%%%%%%%%%%%%%%%%%%%%%%%%%%%%%%%%%

In the last section we showed that the Killing spinor equations
are solved if the supergravity fields satisfy the eqs.\ \dau\  and
\daq\ with the metric ansatz given by \dag.
Following \LOSW, let us consider a simple example $Y=T^6$,
where the intersection form is given by:
\eqn\eaa{
{1 \over 6} C_{IJK} X^I X^J X^K = X^1 X^2 X^3
}
For this example the equations \dau\ become:
\eqn\eab{\eqalign{
& H_1 = e^{-2U} \, (X^2)(X^3) \cr
& H_2 = e^{-2U} \, (X^1)(X^3) \cr
& H_3 = e^{-2U} \, (X^1)(X^2)
}}
and thus:
\eqn\aec{
X^1 = {e^{2U} \over H_1} \quad , \quad
X^2 = {e^{2U} \over H_2} \quad , \quad
X^3 = {e^{2U} \over H_3} \quad , \quad
e^{6U} = H_1 H_2 H_3
}
For the generic case of a ``running volume'' ($\gamma =-4$)
the domain wall metric reads: 
\eqn\aed{
ds^2 = (H_1 H_2 H_3)^{1/3} \Big[ -dt^2 + dx_1^2 + dx_2^2 +dx_3^2 \Big]
+ (H_1 H_2 H_3)^{4/3} dy^2}
and the volume is $\CV = e^{6U} = H_1 H_2 H_3$ (setting $\ell =1$).
In order to understand the domain wall from the M theory
perspective, we can rescale the solution  and obtain for
the string frame and for the K\"ahler class moduli:
\eqn\aef{\eqalign{
& ds_{str}^2 = (H_1 H_2 H_3)^{-1/3} \Big[ -dt^2 + dx_1^2 + dx_2^2 +dx_3^2 \Big]
+ (H_1 H_2 H_3)^{2/3} dy^2 \cr
& t^I = \CV^{1/3} X^I = e^{2U} X^I = {(H_1 H_2 H_3)^{2/3} \over H_I}
}}
In an infinite volume limit we can decompactify this solution
and the 11-d metric becomes
\eqn\aet{
ds_{11}^2 = {1 \over (H_1 H_2 H_3)^{1/3}} \Big[ -dt^2 + dx_1^2
+dx_2^2 + dx_3^2 + (H_2 H_3 d\omega_1 + cycl.) + H_1 H_2 H_3 dy^2 \Big]
}
where $d\omega_{1,2,3}$ are 2-d line elements and this configuration
is an intersection $M5 \times M5 \times M5$ over a common 3-brane.

On the other hand in the fixed volume case, the $X^I$ field are the same,
but since $\gamma =2$  the metric differs
\eqn\kos{
ds^2 = (H_1 H_2 H_3)^{1/3} \Big[ -dt^2 + dx_1^2 + dx_2^2 +dx_3^2 \Big]
+ (H_1 H_2 H_3)^{- 2/3} dy^2}
which yields $AdS_5$ for large $y$.

%%%%%%%%%%%%%%%%%%%%%%%%%%%%%%%%%%%%%%%%%%%%%%%%%%%%%%

\subsec{Discussion of some global aspects}

%%%%%%%%%%%%%%%%%%%%%%%%%%%%%%%%%%%%%%%%%%%%%%%%%%%%%

Solving the local supergravity equations is not enough to describe
domain walls, which are typically gravitational kink solutions that
interpolate between vacua at $y=\pm \infty$. Interesting cases are
interpolating solutions between vacua with different cosmological
constants on both sides, i.e.\ the scalar fields flow between extrema
of the potential. But there are also dilatonic domain walls, where the
potential typically does not allow for isolated extrema and at least
one scalar field ``runs away''. This resembles the linear dilaton
vacua appearing in certain string backgrounds. In the compactified
theory, this run-away behavior signals a strong or weak coupling
region, where the internal volume either diverges or shrinks. This is
exactly the case for the solution that we described, where the volume
of the internal space as described by the volume scalar $\varphi$
diverges for $y \rightarrow + \infty$. On the other hand, the scalars
in the vector multiplets are fixed by the attractor equations \dau\
and become constant asymptotically, fixed only by the flux vector
$\alpha_I$.  Therefore, the Killing spinor equations imply that
asymptotically $\partial_A W =0$ and we reach an extremum of the
superpotential $W(\phi^A)$.

Note, the supergravity solution for the scalars $X^I(y)$ describes a
trajectory on the (curved) moduli space and since it solves the equations
of motion this trajectory is geodesic with radial coordinate $y$ as
affine parameter. This geodesic is fixed if we fix the two endpoints,
i.e.\ two vacua. Let us stress, that it is
not enough to fix only one endpoint, say, at $y=+\infty$, we have also to
choose where the solution should flow at $y=-\infty$, i.e.~on the other side
of the wall.

But what happens if we pass the point $y=0$? Our solution is valid for
all values of $y$ and the point $y=0$ is generically not singular. So,
we have to discuss the continuation to negative values of $y$. Because
the domain wall is an intersection of five-branes, the flux-vector
$\alpha_I$ should change at least the sign while passing the
five-branes. As consequence, the product $\alpha_I y$ remains positive
and we avoid singularities due to zeros of harmonic
functions\foot{Note, a vanishing harmonic function $H_I$ means a
vanishing cycle $X_I$.} at finite values of $y$. A trivial possibility
is to treat both sides symmetrically and therewith identifying both
asymptotic vacua. More interesting, especially {from} the RG-flow
point of view, is to patch together different vacua.  An interesting
case would be a solution interpolating between different vacua of a
given superpotential $W$, but these domain walls are expected to be
singular, because due to the global convexity of the moduli space
\MOR\ the attractor equations \dau\ have only one solution for a given
Kaehler cone \HP\ and different extrema of $W$ have to lie on
disconnected branches of the moduli space. One can also consider a
domain wall describing the flow towards vanishing volume. In this case
the metric develops a singularity where the 4-dimensional world volume
is squeezed to zero size, for examples see \refs{\GPPZZ, \BC, \KA}. 
Let us comment on them in more detail.

The first thing to notice is that we can always approach this singular
point by a proper choice of the vector $q_I$, which fixes the point
$X^I(y=0)$. A vanishing volume of the internal space yields always a
singularity in supergravity solutions, but as we discussed earlier,
quantum corrections or better higher curvature corrections provide a
cut-off for the volume. This lower bound \uhypershift\ was basically
given by the Euler number of the internal manifold and therefore the
regular supergravity solution can allow for at most 4 unbroken
supercharges.  By a simple shift in $y$ we can always arrange that we
reach this ``quantum volume'' at $y=0$ and it is natural to describe
the other side of the wall by the solution with a fixed volume, i.e.\
$\gamma =2$ in the solution described in section 4.1. Therefore in
this regularized supergravity solution, the volume scalar flows from
infinity (infinite volume) towards a lower bound and the scalars in
the vector multiplets extremize on both sides the superpotential,
i.e.\ they flow between fixpoints. Notice, the superpotential does not
need to be the same on both sides, e.g.\ we may change the flux vector
$\alpha_I$ on both sides and/or the intersection form but, due to the
attractor equation combined with the convexity of $\CM$, a given
superpotential $W(\phi^A)$ has a unique extremum, where the flux
vector $\alpha_I$ is parallel to the normal vector $X_I$ (see
Fig.~1). Moreover this type of domain wall provides an interesting
example from AdS/CFT perspective, because the gauge theory couplings
which are dual to the Kaehler class moduli $t^I$ are UV-free, related
on the sugra side to the infinite volume region, and flow in the IR to
a non-trivial conformal fixpoints, where the supergravity solution
becomes $AdS_5$ with fixed scalars.

%%%%%%%%%%%%%%%%%%%%%%%%%%%%%%%%%%%%%%%%

\vskip 1cm
\centerline{\bf Acknowledgements}

We would like to thank K.~\& M.~Becker, J.~Gomis,
C.~Vafa and E.~Witten for useful discussions.
The work of K.B. was supported by a Heisenberg Fellowship of the DFG.
The work of S.G. was supported in part by the Caltech Discovery Fund,
grant RFBR No 98-02-16575 and Russian President's grant No 96-15-96939.

\vskip 1cm

\listrefs
\end